\documentclass[%
 reprint,
 amsmath,amssymb,
 aps,
]{revtex4-2}

\usepackage{graphicx}
\usepackage{dcolumn}
\usepackage{bm}

\begin{document}

\preprint{APS/123-QED}

\title{A Feasible Hybrid Quantum-Assisted Digital Signature for Arbitrary Message Length}

\author{Marta Irene Garc\'ia Cid}
\affiliation{%
 Universidad Polit\'ecnica de Madrid, Campus Montegancedo, Madrid, Spain\\
}%
\affiliation{
 Indra Sistemas S.A., Madrid, Spain
}%

\author{Laura Ortiz Mart\'in}
\affiliation{
 Universidad Polit\'ecnica de Madrid, Campus Montegancedo, Madrid, Spain\\
}%

\author{David Domingo Mart\'in}
\affiliation{%
 Indra Sistemas S.A., Madrid, Spain
}%

\author{Rodrigo Mart\'in S\'anchez-Ledesma}
\affiliation{%
 Indra Sistemas S.A., Madrid, Spain
}%

\author{Juan Pedro Brito M\'endez}
\affiliation{
 Universidad Polit\'ecnica de Madrid, Campus Montegancedo, Madrid, Spain\\
}%

\author{Vicente Mart\'in Ayuso}
\affiliation{
 Universidad Polit\'ecnica de Madrid, Campus Montegancedo, Madrid, Spain\\
}%

\date{\today}

\begin{abstract}
Currently used digital signatures based on asymmetric cryptography will be vulnerable to quantum computers running Shor's algorithm. In this work, we propose a new quantum-assisted digital signature protocol based on symmetric keys generated by QKD, that allows signing and verifying messages in a simple way implementing an integration of currently available classical and quantum technologies. The protocol is described for a three-user scenario composed of one sender and two receivers. In contrast to previous schemes, it is independent of the message length. The security of the protocol has been analyzed, as well as its integrity, authenticity and non-repudiation properties. \\

\end{abstract}

\maketitle


\section{\label{sec:Intro}Introduction}

Cryptography guarantees the confidentiality, authentication, integrity and non-repudiation of transmitted information \cite{DefCrypto}. There are two major answers within cryptography to ensure these four properties: encryption and digital signature (DS). Encrypting communications ensures the confidentiality of the transmitted information between sender and receiver, that is, that no unauthorized user can extract and understand this information. Meanwhile, digital signature aims to guarantee the identity of the sender of a message, as well as the integrity and non-repudiation of the message. The former meaning that the content of the message has not been modified after signing, and the later meaning that the signer can not deny that he has signed the message.\\
Over the years, a large number of DS protocols have been proposed, analyzed and standardized. Examples of digital signature algorithms are Lamport-Diffie's \cite{Lamport-Diffie}, RSA \cite{RSA}, El Gamal \cite{ElGamal} and signatures based on elliptic curves \cite{ECDSA}, among others. Their security is based on one-way functions, i.e. mathematical problems that are easy to calculate in one direction but  extremely hard to calculate in the reverse way. A typical example is prime factoring; while is easy to multiply two numbers, it is very difficult to find the prime factors
of a large number. It is important
to note that there is no demonstration that one way functions exist. These protocols make use of a public-private key system whose security relies in the difficulty of solving such a mathematical problem, i.e. in its computational complexity.\\
The arrival of the quantum computer, has altered the computational complexity panorama. In particular, the ability to implement Shor's or Grover's algorithms, make the DS algorithms vulnerable \cite{Shor,Grover}. \\
To solve this security challenge, two new cryptographic paradigms have been explored in recent years: quantum cryptography \cite{BB84,E91} and post-quantum cryptography (PQC) \cite{Code,Lattice,Poly}. In PQC, cryptographic algorithms based on new mathematical problems are defined, such as lattices, codes, etc \cite{whitepaperETSI}. It should be noted that PQC is purely classical, that is, it is not based on any quantum phenomenon, unlike quantum cryptography. PQC schemes are currently under a standardization process launched by the National Institute of Standards and Technology (NIST) in 2017 \cite{NIST}. Candidates are selected for both Public-key encryption and Key-encapsulation mechanisms (KEM) and for Digital Signature Algorithms (DSA). At the time of writing this work, NIST has started the standardization of the finalist DS algorithms CRYSTALS-DILITHIUM \cite{CRYSTALS-DILITHIUM}, FALCON \cite{FALCON} and SPHINCS+ \cite{SPHINCS+} and, at the same time, a fourth round of the evaluation process has started for the KEM algorithms as well as a Call for Proposals for new DSA. The process is expected to be completed in 2024. A timeline for the phase-out of the current schemes, based on RSA or ECC was also suggested. However, it is important to note that these new algorithms are demonstrated to be resistant only against a quantum computer running Shor's algorithm and it is lamentably unknown what a full-fledged quantum computer might do. It is also important to note that the new algorithms have not been scrutinized so deeply as current algorithms, a fact put in sharp evidence during the NIST process, when algorithms in the last round where unexpectedly broken using personal computers \cite{PQC_roto}. \\ 
On the other hand, quantum cryptography uses cryptographic protocols based on quantum mechanical properties. These protocols do not depend on any mathematical problem, but are based on the laws of physics. Consequently, they are immune to computational attacks and offer a security that does not depend on assumptions on the computational capabilities of the attacker, a fact known as Information Theoretic Security (ITS). The most widely known quantum cryptographic protocols are those used for Quantum Key Distribution (QKD) \cite{BB84,E91}, which allows the generation of a random symmetric key between two users by using the quantum properties of photons. Currently, commercial devices are already available to execute QKD protocols \cite{IDQ,Toshiba}, which have been implemented in long-range networks and whose QKD-generated keys have been employed to encrypt transmitted information \cite{RedMadrid,RedChina}. But quantum cryptography encompasses types of protocols, such as the quantum digital signature (QDS), which is the subject of this work.\\
The first QDS was proposed in 2001 \cite{PrimerQDS} and was based on Lamport's one-way function scheme \cite{Lamport-Diffie}. This signature scheme has been demonstrated ITS but, despite being a robust QDS, its implementation is currently not possible since it requires the use of long-term quantum memories. In 2014, a QDS scheme was proposed that did not require quantum memories for its implementation since the measurement of quantum states is done upon receiving them \cite{QDS-2014}. In this case, the main disadvantages are the vulnerability to coherent forging attacks. In addition, the presence of spies is not allowed in the analysis which means that the use of secure quantum channels must be guaranteed. Two years later, a new scheme was published whose security no longer depends on the availability of secure quantum channels \cite{QDS-2016}, but on secure classical channels in order to carry out a symmetrization step. In this case, the signer receives and measures the quantum states of the receivers, without carrying out error correction and privacy amplification mechanisms corresponding to the key distillation processes, which are used both for the signature and for the securitization of the classical channels. These types of protocols have certain limitations regarding the maximum length of the message that can be signed, i.e. 1 bit, and the number of participants, which decreases the protocol efficiency as they increase. More recently, in 2021 a scheme was proposed where the symmetrization step \cite{QDS-2021} was removed. With this improvement, the scheme achieves greater execution efficiency, but the limitation regarding the length of the signed message remains the same. \\
In this work, we propose a new pseudo-symmetric quantum-assisted digital signature protocol implemented with current technology, which simplifies the previous schemes and allows to sign messages with an arbitrary length. Using keys generated by QKD as the basis of the scheme, the protocol is built taking into account hash functions and current NIST recommendations, thus providing a hybrid system that integrates quantum and classical cryptography. \\
The article is structured as follows, Section \ref{QDS Protocol} explains the new quantum-assisted digital signature protocol. The security of the protocol has been analyzed in Section \ref{Security Proof}, in order to confirm its robustness regarding integrity, authenticity and non-repudiation properties. The article end up with the final conclusions.

\subsection{\label{QDS Protocol}Quantum-assisted digital signature Protocol}
By definition, a digital signature, whether classical or quantum-assisted, must fulfill a series of characteristics \cite{Pastor98}:
\begin{enumerate}
    \item Dependence with the content of the message;
    \item Generation using secret information  from the sender;
    \item Efficiency in the generation and verification processes;
    \item Unforgeable and non-repudiable.
\end{enumerate}

These characteristics ensure the properties of authenticity, integrity and non-repudiation that all digital signatures must guarantee. The quantum-assisted digital signature protocol proposed in this work does indeed meet each of these requirements (see Section \ref{Security Proof}).\\
The general structure of the new pseudo-symmetric quantum-assisted digital signature (Q-DS) protocol is divided in two phases. A \textbf{distribution phase}, in which Alice establishes secret symmetric keys through quantum key distribution (QKD) processes with all possible receivers. After that, the receivers exchange between them random elements of those keys. These QKD-generated keys are the foundation of the security in the protocol, since the keys are ITS. These keys have the security features provided by QKD systems \cite{BB84_Security}. As already mentioned, it is an ITS key that is not vulnerable to brute force attacks nor does it depend on the solution to a complex mathematical problem, plus no one else knows that key but Bob and Alice, which guarantees its confidentiality, since the knowledge that an eavesdropper can have over the key can be made arbitrarily small.
A \textbf{messaging phase}, in which Alice generates the digital signature for a given message and sends it to the recipients who verify the validity of both the message and the signature. This second phase involves completely classical procedures and can be carried out some time after the first phase. However, in order to create the digital signature of the message, it is necessary to rely on the secret symmetric keys from the distribution phase, which have been generated by QKD. This is why, although the digital signature production and verification processes are classic, the full protocol is meant to be quantum-assisted. These two phases of the protocol are described in detail in the following subsections, assuming a scenario with three users, one signer (Alice) and two verifiers (Bob y Charlie).

\subsection{Distribution Phase}
\begin{figure}
\includegraphics[width=0.5\textwidth]{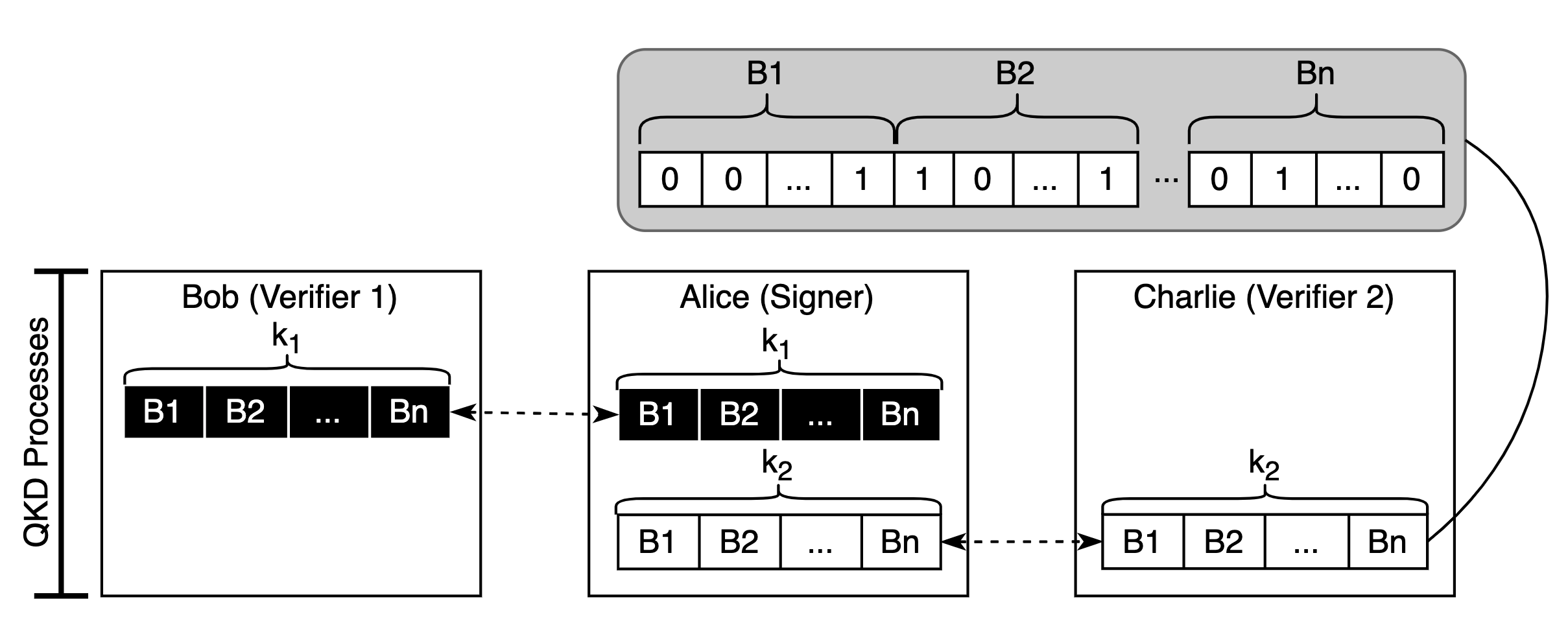}
\caption{\label{QKD-diagram}Distribution phase of the proposed Quantum-assisted Digital Signature scheme. A Quantum key Distribution protocol is carried out between Alice-Bob and Alice-Charlie to generate $k_1$ and $k_2$, respectively. The keys are divided in $n$ blocks for the exchange of random blocks between Bob and Charlie.}
\end{figure}

The protocol begins with the distribution phase in which Alice and Bob carry out a QKD process, for example the well-known BB84 protocol \cite{BB84}. This protocol is executed and the result is a secret symmetric key $k_1$ of length $l$, shared by Alice and Bob.\\
These same steps are carried out between Alice and Charlie, obtaining at the end of the process a secret symmetric key $k_2$ of length $l$.\\
The keys $k_1$ and $k_2$ are divided into $n$ blocks of length $\frac{l}{n}$, as shown in Figure \ref{QKD-diagram}, and stored. Each block has a fixed labeling $B=\{B1, B2, ..., Bn\}$ which is known by all the users. Once the generation of symmetric keys between pairs is finished, Bob and Charlie carry out a process of exchanging random blocks of key. For that, given $S_n$ as the group of permutations of n elements so that $|S_n|=n!$, we denote the random  permutation of n elements $\gamma_j \in S_n$, with $j=\{b,c\}$ as:
\begin{equation}
\gamma_j =
\begin{pmatrix}
1 & 2 & \cdots & n\\
a_1 & a_2 & \cdots & a_n
\end{pmatrix}
, \gamma_j (i) = a_i, a_i \in \{1, 2, ..., n\}
\end{equation}
Bob sends to Charlie through a classic authenticated and encrypted channel (for example with other keys generated by QKD) the subset of elements $k'_1$ corresponding to the first $n/2$ indices of $B$ associated to $k_1$ after applying $\gamma_b$. For its part, Charlie sends to Bob the subset of elements $k'_2$ corresponding to the first $n/2$ indices of $B$ associated $k_2$ after applying $\gamma_c$. At no point does Alice know which blocks Bob and Charlie have exchanged. This is a necessary condition so that Alice cannot repudiate the authorship of the message, as we will see later.\\
The distribution phase does not have to be carried out immediately before the messaging phase, but the users may have generated and stored several keys beforehand to be used later, as long as the security requirements for the maximum storage time of the keys defined by the users are fulfilled.

\subsection{Messaging Phase}
\begin{figure*}
\includegraphics[width=\textwidth]{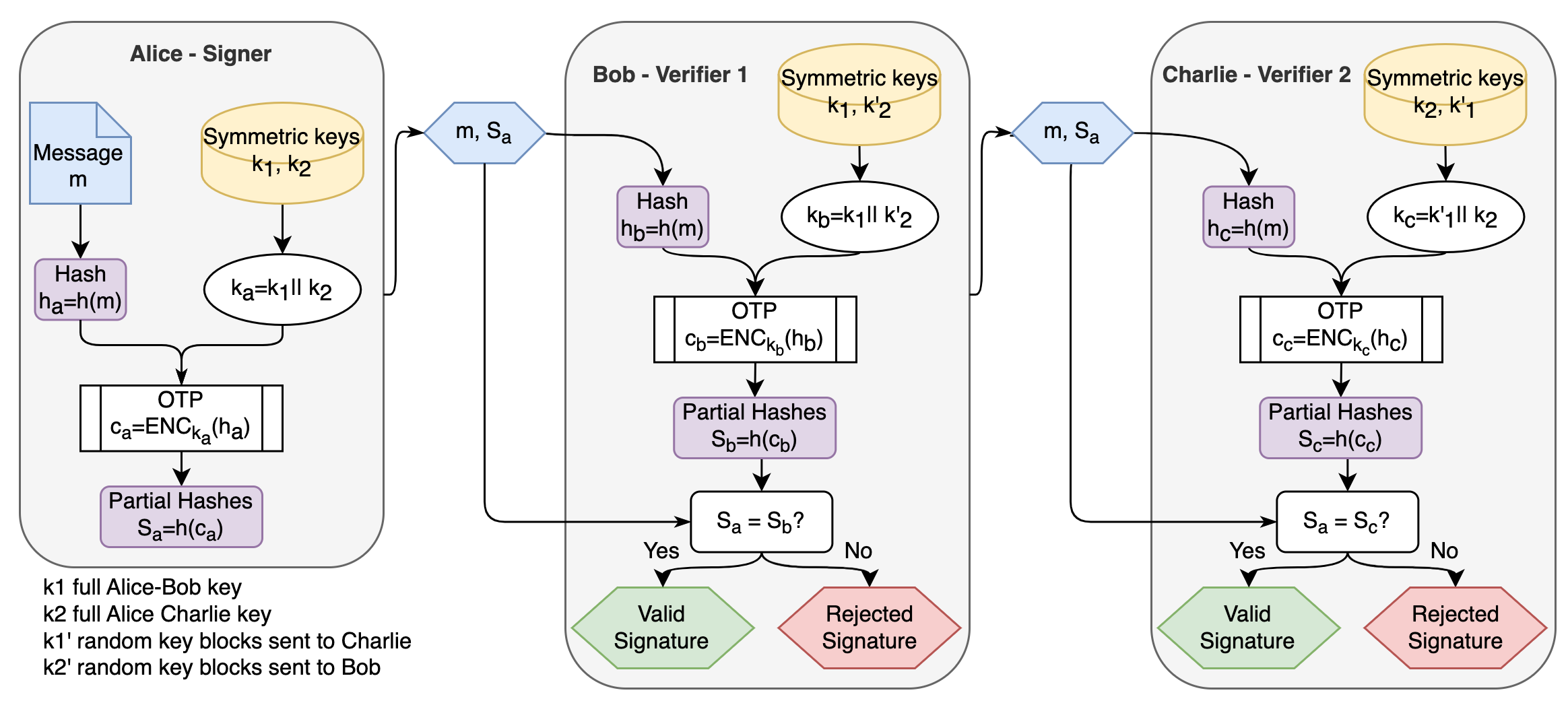}
\caption{\label{Q-DS-Diagram}Messaging phase of the proposed Quantum-assisted digital signature scheme. Alice generates the signature $S_a$ associated to a message $m$ and sends both to Bob who verifies their validity. Then, the message and the signature is forwarded to Charlie who also verifies them. OTP=One-Time Pad.}
\end{figure*}

Then, the messaging phase begins, following the flow shown in Figure \ref{Q-DS-Diagram}. Alice, the signer, wants to send Bob, the first verifier, a message ($m$) of any length and its associated signature ($S_a$). To do so, Alice generates a combined key $k_a$ of length $2l$ by concatenating the secret symmetric key shared with Bob to the secret symmetric key shared with Charlie ($k_a = k_1||k_2$). \\

The novelty is using a hash function by which takes as input a text of any size ($D$) and returns a fixed-size digest equal to $d$. Alice calculates the hash of the message as $h_a = h(m)$. These functions have the property of being one-way functions or preimage resistant since knowing the value of $h_a$ makes it computationally infeasible to extract the value of $m$, under certain conditions as explained in Section \ref{Security Proof}. \\
Then, Alice encrypts $h_a$ with the combined secret key $k_a$ using a One-Time Pad (OTP) encryption function, whereby an element-by-element XOR is made between $h_a$ and $k_a$, obtaining $c_a = ENC_{k_a}(h_a)$. In order to carry out this step $h_a$ and $k_a$ must have the same length, so $2l=d$. \\
At this point, it is important to remember that $k_1$ and $k_2$ were divided into $n$ blocks each (see Figure \ref{QKD-diagram}), so $k_a$ has $2n$ blocks. And, in turn, $c_a$ will also be made up of $2n$ blocks. As a last step, Alice generates individual hash for each of the blocks of $c_a$, thus obtaining the signature $S_a$ of length $2l\cdot 2n = 4nl$, associated with her message.\\
Alice then sends to Bob the ($m$, $S_a$) tuple. Once the information is received, Bob starts the signature verification process, which consists in performing the same steps as those performed by Alice for the generation of $S_a$. Note that in this case, Alice sent the message without encryption, because the objective of this protocol is not confidentiality of the message but authenticity, integrity and non-repudiation. Thus, Bob makes the concatenation of the secret symmetric key shared with Alice and the known blocks of Charlie's key ($k_b = k_1||k'_2$). By using the same hash function as Alice, Bob calculates the hash of the message as $h_b = h(m)$ and encrypts $h_b$ with $k_b$ also using an OTP cipher, $c_b = ENC_{k_b}(h_b)$. Finally, Bob generates the hash of each of the blocks of $c_b$, taking into account that he does not know half of the blocks in $c_b$ that have the same labeling as the unknown blocks in $k'_2$. Doing this he obtains $S_b$ and compares it with the received $S_a$ in such a way that, if all the known elements match $S_a \approx S_b$, being the number of coincidences above a predefined verification threshold $V_B$ (see Section \ref{Sec-Repudiation}), Bob accepts the message and the signature as genuine. However, if the number of coincidences is below $V_B$, $S_a \neq S_b$ and Bob rejects the validity of the message and/or the signature. The signature $S_a$ is approximately equal to $S_b$ but not strictly equal because the information that each of the users has about the keys, $\{k_a, k_b\}$, is different due to the partial exchange of blocks. \\
For completing the sending chain, Bob forwards the tuple ($m$, $S_a$) to Charlie, who performs the same signature verification procedure as Bob:

\begin{enumerate}
    \item Concatenation of the key shared with Alice and the known blocks of Bob's key ($k_c = k'_1||k_2$);
    \item Calculation of the hash of the message as $h_c = h(m)$;
    \item OTP Encryption of $h_c$ with $k_c$ getting $c_c = ENC_{k_c}(h_c)$;
    \item Generation of the hash of each of the blocks of $c_c$ obtaining $S_c$;
    \item Comparison of $S_c$ with $S_a$, for a verification threshold $V_C$:
    \begin{itemize}
        \item Acceptance of the message and the signature as genuine if $S_a \approx S_c$; 
        \item Rejection of the validity of the message and/or the signature if $S_a \neq S_c$. 
    \end{itemize}
\end{enumerate}

\section{Security proof} \label{Security Proof}
In this section, the security of the proposed Q-DS protocol is analyzed from different perspectives. For this aim, malicious users will be introduced to verify that the scheme is robust against attacks on message integrity, forgery attempts and message repudiation. The section ends with a study of the security strength of the cryptographic functions that make up the scheme.

\subsection{Integrity}
We are going to validate the robustness of the scheme against the integrity of the message that is sent, that is, if $m$ is modified along the way, this will be detected and the protocol aborted. The entire process is collected in Table \ref{Integrity}, which sequentially shows the steps followed by each of the users. To do so, we assume the scenario described in Section \ref{QDS Protocol} but, in this case, Bob is a malicious user. \\
It begins with Alice generating the quantum-assisted digital signature with the usual steps indicated in the second column of the table. Bob receives $(m, S_a)$ from Alice and verifies and accepts the received signature as shown in the third column of Table \ref{Integrity}. After that, he modifies the message $m \rightarrow M$, but keeping the signature generated by Alice $S_a$. Bob sends to Charlie the tuple $(M, S_a)$, with the modified message. Upon receiving it, Charlie performs the verification process as shown in the fourth column of the table and at the end of the process detects that $S_c \neq S_a$, so he rejects the signature and the message. 

\begin{table*}[]
\caption{\label{Integrity}Q-DS protocol with a message integrity attack attempt. Alice generates $(m, S_a)$ and Bob verifies and accepts it. Bob sends Charlie the modified message $M$ along with Alice's original signature. Charlie detects that $S_c \neq S_a$ and rejects the signature and the message. The elements of the chain that are approximately equal but not strictly equal is because the information that each of the users has about the keys is different due to the partial exchange of blocks.}
    \begin{tabular}{|l|c|c|c|c|}\hline
         & Alice & Bob & Charlie & Parameters Comparison \\ \hline
         Initial data & $m, k_1, k_2$ & $M, k_1, k'_2$ & $k'_1, k_2$ & $-$ \\ \hline
         Step 1 & $k_a=k_1||k_2$ & $k_b=k_1||k'_2$ & $k_c=k'_1||k_2$ & $k_a \approx k_b \approx k_c$ \\ \hline
         Step 2 & $h_a=h(m)$ & $h_b=h(m)$ & $h_c=h(M)$ & $h_a = h_b \neq h_c$ \\ \hline
         Step 3 & $c_a = ENC_{k_a}(h_a)$ & $c_b = ENC_{k_b}(h_b)$ & $c_c = ENC_{k_c}(h_c)$ & $c_a \approx c_b \neq c_c$ \\ \hline
         Step 4 & $S_a=h_p(c_a)$ & $S_b=h_p(c_b)$ & $S_c=h_p(c_c)$ & $S_a \approx S_b \neq S_c$ \\ \hline
         Verification & $-$ & Accepted & Rejected & $-$ \\ \hline
    \end{tabular}
\end{table*} 

The fifth column in Table \ref{Integrity} compares step by step the parameters generated by each of the users during their turn and indicates the inequalities produced by the malicious Bob that lead to the final rejection of $(m, S_a)$. Note that in this case, Charlie cannot distinguish whether the reason for the negative verification test was an attack on the integrity of the message or a forgery attempt of the signature.\\
In a scenario in which all users are honest, when doing the parameter comparison all of them would be strictly equal ($h_x$) or approximately equal ($k_x, c_x, S_x$), respectively (due to the unknown elements of $k_1$ and $k_2$ by Charlie and Bob). But, in this case, modifying the message causes the first alteration in the calculation of the message hash. Since the calculation of $S_a$ and $S_c$ depends on the value of $c_a$ and $c_c$, respectively, and these depend on the value of $h_a$ and $h_c$, the alteration produced by the change of message is propagated along the entire chain, triggering an error in Charlie's verification test.\\
In order to succeed in an attack on the integrity of the message, Bob would have to be able to find a second preimage of the hash value that is, from the given message digest of $m$, $h(m)$, finding a different input $M$ that provides the same hash value $h(M)=h(m)$. A second preimage attack can also be prevented applying certain conditions (see section \ref{Security Strength}).

\subsection{Forgery}
\label{sec-forgery}
In this scenario, Bob is once again a malicious user. As in the integrity test, Table \ref{Forgery} shows Bob's attempted forgery of Alice's signature. As before, the second column shows the generation of the signature $S_a$ by Alice and the third column indicates the steps taken by Bob for the verification and subsequent acceptance of the message and the signature. 

\begin{table*}[]
    \caption{\label{Forgery}Q-DS protocol with a message and signature forgery attack attempt. Alice generates $(m, S_a)$ and Bob verifies and accepts it. Bob forges Alice's signature and sends Charlie the modified message $M$ and the forged signature $S_f$. Charlie detects that $S_c \neq S_f$ and rejects the signature and the message.}
    \centering
    \begin{tabular}{|l|c|c|c|c|c|}\hline
         & Alice & \multicolumn{2}{c|}{Bob} & Charlie & Param. Comp. \\ \hline
         & & Verification & Forgery & & \\ \hline
         Init. data & $m, k_1, k_2$ & \multicolumn{2}{c|}{$M, k_1, k'_2, K$} & $k'_1, k_2$ & $-$ \\ \hline
         Step 1 & $k_a=k_1||k_2$ & $k_b=k_1||k'_2$ & $k_f=k_1||K$ & $k_c=k'_1||k_2$ & $k_a \approx k_b \approx k_c \neq k_f$ \\ \hline
         Step 2 & $h_a=h(m)$ & $h_b=h(m)$ & $h_f=h(M)$ & $h_c=h(M)$ & $h_a = h_b \neq h_f = h_c$ \\ \hline
         Step 3 & $c_a = ENC_{k_a}(h_a)$ & $c_b = ENC_{k_b}(h_b)$ & $c_f = ENC_{k_f}(h_f)$ & $c_c = ENC_{k_c}(h_c)$ & $c_a \approx c_b \neq c_c \neq c_f$ \\ \hline
         Step 4 & $S_a=h_p(c_a)$ & $S_b=h_p(c_b)$ & $S_f=h_p(c_f)$ & $S_c=h_p(c_c)$ & $S_a \approx S_b \neq S_c \neq S_f$ \\ \hline
         Verif. & $-$ & Accepted & $-$ & Rejected & $-$ \\ \hline
    \end{tabular}
\end{table*} 

In this case, the fourth column shows how Bob generates a fake digital signature $S_f$, associated to the tampered message $M$. As it can be seen, the signature generation process is the same as Alice's, except that Bob does not know the $50\%$ of Charlie's key $k_2$, so the key $K \neq k_2 $ used to generate $k_f$ will produce the alteration shown in the comparison of parameters in the sixth column. This alteration will propagate through all the steps of the Charlie verification test and will result in $S_f \neq S_c$, and the message and signature rejected. 

In order to successfully forge the signature, Bob would have to be able to guess the unknown elements of $k_2$. To do this, he could:

\begin{enumerate}
    \item Guess the unknown bits of $k_2$. Due to the randomness of the QKD-generated key, he has a $50\%$ chance of matching the value of each bit. If he has to guess $0.5l$ bits, where $l$ is the length of $k_2$, the probability that all his guesses are correct is $\frac{1}{2^{l/2}}$. 
    For a $k_2$ length of 112 bits then, $P_{guess}\approx 10^{-17}$ and for 256 bits, $P_{guess}\approx 3\cdot 10^{-39}$. Therefore, the larger the original key length, the smaller Bob's probability of success.
    \item Extract the unknown elements of $k_2$ from the signature $S_a$. If block hashes were not done in the last step of signature generation, Bob could compare $c_c$ and $c_a$, and extract the remaining $k_2$ elements simply by undoing the OTP cipher. But, block hashes of certain size prevent him from accessing the full value of $c_a$ and therefore the keys from being extracted. 
    \item Find a preimage, that is, given a hash value $H$, identifying the input message $m$ that provides $h(m)=H$. These kind of attacks are mainly done by brute force but can be prevented by taking into account a series of security parameters and requirements in the choice of hash functions and the input length, as it is explained in section \ref{Security Strength}.
\end{enumerate}

\subsection{Repudiation}
\label{Sec-Repudiation}
Finally, we assume a scenario where Alice is the malicious user whose goal is to generate a signature that Bob accepts but Charlie rejects, thus denying authorship of her message.\\
Once again, the steps followed by each of the users in this scenario is shown in Table \ref{Repudiation}. Before starting the process, Alice is aware that Bob only knows half of Charlie's key $k_2$, but she has no clue about which elements Bob knows. For Bob to accept the signature and Charlie to reject it, Alice is going to concatenate Bob's and Charlie's keys as usual, but introducing some errors in random bits of $k_2$, so that $k_2 \rightarrow K$. For the strategy to be successful, Alice has to ensure that $100\%$ of the errors introduced are in the elements of $k_2$ that Bob does not know about. Otherwise, Bob detects that $S_a \neq S_b$, as shown in the third column of the table, rejects the validity of the signature and aborts the protocol.

\begin{table*}[]
    \caption{\label{Repudiation}Q-DS protocol with a repudiation attempt. Alice generates $(m, S_a)$ introducing some errors in $k_2$ and sends it to Bob. Bob performs the verification test and detects that $S_b \neq S_a$, so he rejects the signature and the message and aborts the protocol.}
    \centering
    \begin{tabular}{|l|c|c|c|c|}\hline
         & Alice & Bob & Charlie & Parameters Comparison \\ \hline
         Initial data & $m, k_1, k_2, K$ & $k_1, k'_2$ & $k'_1, k_2$ & $-$ \\ \hline
         Step 1 & $k_a=k_1||K$ & $k_b=k_1||k'_2$ & $-$ & $k_a \neq k_b$ \\ \hline
         Step 2 & $h_a=h(m)$ & $h_b=h(m)$ & $-$ & $h_a = h_b$ \\ \hline
         Step 3 & $c_a = ENC_{k_a}(h_a)$ & $c_b = ENC_{k_b}(h_b)$ & $-$ & $c_a \neq c_b$ \\ \hline
         Step 4 & $S_a=h_p(c_a)$ & $S_b=h_p(c_b)$ & $-$ & $S_a \neq S_b$ \\ \hline
         Verification & $-$ & Rejected & Aborted & $-$ \\ \hline
    \end{tabular}
\end{table*} 

The probability for Alice to introduce all the errors in the unknown blocks by Bob depends on the total number of blocks and the number of blocks where introduce errors. If she introduces errors in 1 of the blocks ($e=1$) the probability of success will be $\frac{n/2}{n}=0.5$, which is independent of $n$, the number of blocks. The probability that Alice introduces errors in the blocks unknown to Bob is given by:

\begin{equation}
P_{rep}= \prod_{i=0}^{e-1}\frac{n-2i}{2(n-i)} 
\end{equation}

As an example, for $e=7$, for a scheme of 32 blocks, the probability that Alice will succeed in making a repudiation attack is $0.34\%$. Then, the more blocks with errors Alice introduces, the lower the probability of success. Based on that, Bob and Charlie can define thresholds of permissible errors or verification thresholds $V_B$ and $V_C$, above which $S_a \neq S_b,S_c$, forcing Alice to introduce a greater number of error and, thus, decreasing her probability of success.

\subsection{Security Strength} \label{Security Strength}
So far we have analyzed the security of our proposed quantum-assisted digital signature scheme against attacks on the integrity of the signed message, falsification of the identity of the signer and attempts to repudiate the message by the author. In this final part, we are going to analyze the security strength of the scheme in terms of its cryptographic functions. We ensure that the protocol is as secure as the most vulnerable of its elements: the secret keys, the hash function used on the message, the OTP encryption and the hash function used on the blocks (partial hashes) - in case it differs from initial hash function.\\
Both the QKD-generated keys and the OTP encryption are classified, under certain conditions, as Information-Theoretic Secure (ITS) \cite{Renner05,Shannon49} and, therefore, conform to perfect secrecy as far as each key is only used once. Therefore, the most vulnerable elements are the hash functions. A priori, the hash functions applied to the message and to the blocks are considered secure as long as one of the approved hash functions defined in the NIST FIPS 180-4 \cite{NIST180-4} or NIST FIPS 202 \cite{NIST-SHA3} standards is used (see Table \ref{Security_Parameters}), with the exception of the SHA-1 function, which is not longer recommended by the NIST. \\

\begin{table}[]
    \caption{\label{Security_Parameters}Strength of NIST-approved hash functions \cite{NIST-SHA3,NIST800-107}. CR=Collision Resistance, PR=Preimage Resistance, 2PR=Second Preimage Resistance. }
    \centering
    \begin{tabular}{|l|c|c|c|}\hline
         & CR (bits) & PR (bits) & 2PR (bits) \\ \hline
         SHA2-224 & 112 & 224 & 201-224  \\ \hline
         SHA2-256 & 128 & 256 & 201-256  \\ \hline
         SHA2-384 & 192 & 384 & 384  \\ \hline
         SHA2-512 & 256 & 512 & 394-512  \\ \hline \hline
         SHA3-224 & 112 & 224 & 224  \\ \hline
         SHA3-256 & 128 & 256 & 256  \\ \hline
         SHA3-384 & 192 & 384 & 384  \\ \hline
         SHA3-512 & 256 & 512 & 512  \\ \hline \hline
         SHAKE-128 & $min(\delta/2, 128)$ & $\geq min(\delta, 128)$ & $min(\delta, 128)$  \\ \hline
         SHAKE-256 & $min(\delta/2, 256)$ & $\geq min(\delta, 256)$ & $min(\delta, 256)$  \\ \hline
    \end{tabular}
\end{table} 

Table \ref{Security_Parameters} lists four of the NIST-approved hash functions from the SHA-2 family and from the SHA-3 family and their strength measured in bits against collisions (CR), preimage (PR), and second preimage (2PR) \cite{NIST-SHA3,NIST800-107}. In section \ref{sec-forgery}, we saw the difference between preimage and second preimage attacks. For its part, a collision attack means finding any two messages $m$ and $M$, being $M \neq m$, such that $h(M)=h(m)$. The difference between collision and second preimage is that in the later the malicious user can be unaware of the content of $m$ if, for example, if it is encrypted. The security strength of a hash function is determined by the lowest of the CR, PR and 2PR strengths. Our Q-DS protocol needs Collision, Preimage and Second preimage resistance, so the security strength is set by CR for all SHA functions. As it can be seen, the functions with the greatest security strength are SHA-512 and SHA-384. 

\subsubsection{Collision resistance.}
The collision resistance depends directly on the space of possible inputs $I$ which has a size $l_i$ and the space of possible output hash values $O$ of size $l_o$, and can be calculated as follows \cite{Calculo_prob}:
\begin{equation}
P_{col} = 1 - exp\left[-\frac{(2^x+1)^2}{2(2^k+1-2^x)}\right]
\end{equation}
Where $x$ is the length in bits of the input message which defines the size of $I$ as $l_i = 2^x$, $k$ is the length of the message digest and defines the size of $O$ as $l_o = 2^k$. For the SHA2-224 function $k=224$ bits, for SHA2-256 function $k=256$ bits, etc. Above the thresholds defined in Table \ref{Security_Parameters}, the probability that there are any two messages $m\neq M$ that produce a collision is different from zero. As an example, the amount of work needed to find a collision for SHA2-256 is $2^{128}$.

\subsubsection{Second preimage resistance.}
As we have seen above, for Bob to successfully carry out a message integrity attack he has to be able to find another $M$ message such that $h(M)=h(m)$.
The calculation of second preimage resistance for SHA2-224, SHA2-256 and SHA2-512 functions, collected in Table \ref{Security_Parameters}, is given by \cite{NIST800-107}:
\begin{equation}
2PR = d - log_2\left(D/B\right)
\end{equation}
Where $d$ is the size of the hash output, $D$ is the size of the input message in bits 
and $B$ is the input block size of the function. 
In the case of SHA2-384, the security strength does not depend on the size of the input message \cite{NIST800-107}, so its resistance to second preimage is given by $d=384$. This value allows us to obtain the amount of work required to find a second preimage, which is $2^{384}$.

\subsubsection{Preimage resistance.}
The most favorable alternative for Bob to forge the signature is to try to find the preimage of each of the unknown key blocks. These types of attacks can be carried out by brute force if the input block size is small enough. As an example, if the block size is 4 bits, it would only take $2^4$ tries to find the value of the block. It is also possible to attack using more efficient methods such as rainbow tables \cite{Rainbow_tables}. \\ 
For this reason, we have to find the optimal balance between having as many blocks as possible to reduce $P_{rep}$, and having the size of each block large enough to be secure against a dictionary attack. In this sense, the use of an extendable-output function (XOF) SHAKE from the SHA-3 family is proposed (see Table \ref{Security_Parameters}), prior to the key-blocks exchange between Bob and Charlie during the distribution phase. This type of functions allows, given an input and a variable $\delta$ parameter, to securely extend the length of the input up to $\delta$ bits \cite{NIST-SHA3}. If Alice, Bob and Charlie carry out this operation in $k_1$ and $k_2$, prior to the block division, they can increase $l\rightarrow \delta$. To maintain the requirement that $2l=d$, this operation can be done directly on the message, modifying its size from $D\rightarrow 2\delta$. As a consequence, the security against dictionary attacks is improved and, furthermore, Bob and Charlie do not exchange plane key blocks but the blocks resulting from the XOFs, which increases the privacy of $k_1$ and $k_2$.

From this analysis we obtain as an example that, if SHAKE-256$(m, \delta=2048)$ is implemented to generate the message hash in the Q-DS protocol, it gives $CR=256$, $PR=256$ and $2PR=256$. We suppose that $k1$ and $k2$ are $l=256$. A SHAKE-256 is applied to $k1$ and $k2$, with $\delta = 1024$. For $n=32$ blocks, the size of each block is $64$ bits. Given these values, the amount of work needed to find a preimage is $2^{64}$. According to these values, if Bob sets a verification thresholds of $V_B=25\%$ (maximum number of blocks known to Bob with errors), then the probability of Alice succeeding in a repudiation attack is $P_{rep}=0.05$.  

Finally, when choosing a proper hash function for the blocks, it could be interesting to consider the SHA2-256 function due to performance reasons, since it already is a function supported by Intel's native instructions \cite{Intel}. 

We conclude the security analysis of the protocol with a brief mention of Denial of Service (DoS) attacks. Any malicious user could simply send wrong information about their key shared with Alice to the rest of users, causing them to reject the signature. Dealing with this issue involves increasing security thresholds, which leads to increasing some of the probabilities of the security analysis. 

\section{Conclusions}
The quantum-assisted digital signature protocol proposed in this article avoids the use of currently employed public-key cryptosystems that are considered potentially vulnerable in a quantum scenario, replacing the key exchange with ITS QKD-generated keys for which commercial devices are available. The use of QKD not only removes the vulnerability of public-private keys, but also, the knowledge that an eavesdropper can have over the key can be made arbitrarily small.\\
Unlike the QDS protocols published to date, the new scheme presented allows to sign messages with an arbitrary length thanks to the use of hash functions, simplifying the previous schemes. This is accomplished by taking into account NIST-recommended hash functions from the SHA-2 and SHA-3 families, thus generating a hybrid system that integrates quantum and classical cryptography implementable with current technology. In addition, a division by blocks of the string and subsequent hashing of the blocks is used to prevent Bob from recovering the unknown elements of Charlie's key and forging the signature. \\
The protocol has been evaluated against attacks on the integrity of the message, forgery of the signature and attempted repudiation by the signer. The first is immediately detected throughout the verification process by the propagation of errors that the modification of the message entails in the initial message digest. In case of a second preimage attack, the minimum amount of work needed is $2^{256}$ in case of using SHAKE-256 for $\delta > 256$. Forgery attacks can be carried out by finding a preimage through brute force or using sophisticated methods as rainbow tables. To avoid them, the implementation of a XOF in the distribution phase allows us to securely extend the length of the blocks before generating the digest of each of them with a SHA2 function, being able to increase the security against dictionary attacks. To prevent a repudiation attack, a partial key exchange step between verifiers is included. Security against repudiation depends on the number of blocks defined, the maximum threshold of permissible errors set by Bob and Charlie independently, and the number of blocks with errors. The number of blocks is required to be as large as possible, thus reducing $P_{rep}$.\\
All these values and the chosen hash functions could vary depending on the use cases and the degree of security required in each of them, providing a Q-DS protocol implementable with the currently available technologies.

\begin{acknowledgments}
This project has received funding from Indra Sistemas S.A; EIT Digital co-funded by European Institute of Innovation and Technology (EIT), a body of the European Union; and MCIN with funding from European Union NextGenerationEU (PRTR-C17.I1) and funding from the Comunidad de Madrid. Programa de Acciones Complementarias. Madrid Quantum.
\end{acknowledgments}

\bibliography{apssamp}

\end{document}